\documentclass[%
 reprint,
 amsmath,amssymb,
 aps,
]{revtex4-1}
\usepackage[usenames,dvipsnames]{xcolor}
\usepackage{amsfonts}
\usepackage{amsmath,amsthm,amssymb,dsfont,etoolbox}
\usepackage{enumerate}
\usepackage{graphicx}	
\usepackage{url}
\usepackage{todonotes}
\usepackage{bbm}

\usepackage{algorithm}
\usepackage{algpseudocode}

\usepackage{tikz}
\usetikzlibrary{chains}
\usetikzlibrary{fit}
\usepackage{pgflibraryarrows}		
\usepackage{pgflibrarysnakes}		
\usepackage{qcircuit}

\usepackage{makecell}

\usepackage{epsfig}
\usetikzlibrary{shapes.symbols,patterns} 
\usepackage{pgfplots}

\usepackage{hyperref}
\hypersetup{colorlinks=true,citecolor=blue,linkcolor=blue,filecolor=blue,urlcolor=blue,breaklinks=true}

\usepackage{nicefrac}
\usepackage{mathtools}

\usepackage{multirow}

\theoremstyle{plain}

\theoremstyle{definition}

\newcommand*{\ee}{\mathrm{e}}

\newcommand*{\cA}{\mathcal{A}}

\newcommand*{\cE}{\mathcal{E}}
\newcommand*{\cF}{\mathcal{F}}

\newcommand*{\cI}{\mathcal{I}}

\newcommand*{\cN}{\mathcal{N}}

\newcommand*{\cT}{\mathcal{T}}

\newcommand*{\cX}{\mathcal{X}}

\newcommand*{\cZ}{\mathcal{Z}}

\newcommand*{\N}{\mathbb{N}}

\newcommand*{\eps}{\varepsilon}
\newcommand*{\diag}{\mathrm{diag}}

\newcommand*{\poly}{\mathrm{poly}}
\newcommand*{\polylog}{\mathrm{polylog}}

\newcommand*{\ket}[1]{| #1 \rangle}
\newcommand*{\bra}[1]{\langle #1 |}

\newcommand{\proj}[1]{|#1\rangle\!\langle #1|}

\newcommand*{\CNOT}{\mathrm{CNOT}}
\newcommand*{\cO}{O}

\newcommand*{\ci}{\mathrm{i}} 

\newcommand{\norm}[1]{\left\lVert#1\right\rVert}
\newcommand{\abs}[1]{\left\lvert#1\right\rvert}

\newcommand{\calS}{{\cal S}}
\newcommand{\calF}{{\cal F}}
\newcommand{\calG}{{\cal G}}

\newcommand{\la}{\langle}
\newcommand{\ra}{\rangle}

\usepackage{float}

\AtBeginEnvironment{pmatrix}{\setlength{\arraycolsep}{0.8pt}}  

\begin{document}
\title{Error mitigation for universal gates on encoded qubits}

\author{Christophe Piveteau}
\affiliation{IBM Quantum, IBM Research -- Zurich, Switzerland}
\affiliation{Institute for Theoretical Physics, ETH Zurich, Switzerland}
\author{David Sutter}
\affiliation{IBM Quantum, IBM Research -- Zurich, Switzerland}
\author{Sergey Bravyi}
\affiliation{IBM Quantum, IBM T.J.~Watson Research Center, Yorktown Heights, US}
\author{Jay M.~Gambetta}
\affiliation{IBM Quantum, IBM T.J.~Watson Research Center, Yorktown Heights, US}
\author{Kristan Temme}
\affiliation{IBM Quantum, IBM T.J.~Watson Research Center, Yorktown Heights, US}

\begin{abstract}
The Eastin-Knill theorem states that no quantum error correcting code can have a universal set of transversal gates. For CSS codes that can implement Clifford gates  transversally it suffices to provide one additional non-Clifford gate, such as the $T$-gate, to achieve universality. Common methods to implement fault-tolerant $T$-gates like magic state distillation generate a significant hardware overhead that will likely prevent their practical usage in the near-term future.  Recently methods have been developed to mitigate the effect of noise  in shallow quantum circuits that are not protected by error correction. Error mitigation methods require no additional hardware resources but suffer from a bad asymptotic scaling and apply only to a restricted class of quantum algorithms. In this work, we combine both approaches and show how to implement encoded Clifford+$T$ circuits where Clifford gates are protected from noise by error correction while errors introduced by noisy encoded $T$-gates are mitigated using  the quasi-probability method. As a result, Clifford+$T$ circuits with a number of $T$-gates inversely proportional to the physical noise rate can be implemented on small error-corrected devices without magic state distillation. We argue that such circuits can be out of reach for state-of-the-art classical simulation algorithms.
\end{abstract}
\maketitle
\textit{Introduction.}---
The universally accepted approach to remedy effects of noise and decoherence in quantum computing is the use of quantum error correction~\cite{shor1995scheme,steane1996error,calderbank1996good}. The celebrated \emph{threshold theorem} guarantees that errors in a quantum computation can be suppressed efficiently to any level for arbitrarily long circuits, if the noise rate of the physical gates is below some \emph{constant} threshold value~\cite{AB08}. The low noise thresholds and the polylogarithmic qubit overhead needed to implement error correction protocols are very demanding on the quantum hardware. However, the by far largest overhead for quantum algorithms with current error correcting codes stems from the implementation of a fault-tolerant universal gate set. 

Clifford gates are typically easier to realize fault-tolerantly than non-Clifford gates for many commonly studied codes. In fact, for many CSS codes they can be implemented transversally.
Transversal gates have a manageable overhead, because by definition these gates are fault-tolerant as they do not spread errors within code blocks. The Eastin-Knill theorem proves that no error correcting code can transversally implement a universal set of gates~\cite{EK09}. 
To obtain a universal gate set a single further non-Clifford gate is needed. The $T$-gate is commonly added for this purpose.

\begin{figure}[!htb]
\centering
\begin{tikzpicture}
\node at (0,0) {
\Qcircuit @C=1em @R=.7em {
&   \targ  & \meter\cwx[1]  \\
& \ctrl{-1} & \gate{SX} &\qw \\
}
};
\node at (-1.4,0.35) {$\ket{\psi}$};
\node at (-1.5,-0.35) {$\ket{\pi/4}$};
\node at (1.5,-0.35) {$T\ket{\psi}$};
\end{tikzpicture}     
    \caption{Implementation of a $T$-gate via a magic state $\ket{\pi/4}$ and Clifford operations. The $S$ and $X$ gates are only performed if the measurement outcome is $1$.}
    \label{fig_Tgate}
\end{figure}
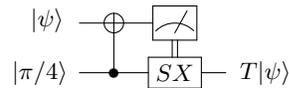

A common approach to implement the $T$-gate fault-tolerantly is the use of reliable encoded magic states~\cite{BK05}. These states can be prepared by a technique called \emph{magic state distillation} and every encoded magic state can be used to perform a $T$-gate using only Clifford operations as shown in Figure~\ref{fig_Tgate}. The distillation of magic states is responsible for the largest overhead in a fault-tolerant computing system~\footnote{Recent results such as~{\cite{Litinski2019magicstate}} show how to reduce the memory footprint of a magic state factory by almost one order of magnitude with roughly the same (or even smaller) space-time volume as a logical $\CNOT$. However, these results only apply to large-distance codes $(d\geq 25)$ and we would need at least $4000$ qubits to get a single $T$-gate.}.

While the hardware requirements to implement quantum algorithms fault-tolerantly have not been met yet, steady progress in the development of quantum hardware has been made~\cite{jurcevic2021demonstration,8936946,bruzewicz2019trapped}. This has given rise to the question: Are there computational tasks that could be implemented without the use of quantum error correction? This question has motivated a set of proposals~\cite{peruzzo2014variational,khatri2019quantum,havlivcek2019supervised,schuld2019quantum,mcardle10variational,mitarai2020theory} that only ask for the implementation of shallow quantum circuits and the estimation of expectation values. Early experimental implementations~\cite{o2016scalable,kandala2017,kandala2019error,havlivcek2019supervised,peters2021machine} of such proposals have shown that the effect of decoherence on the result is non-negligible, despite the use of the restrictive computational models and good noise levels. It has become clear that methods to remove the noise-induced bias from the expectation values need to be applied. These methods are referred to as \emph{quantum error mitigation}~\cite{Temme2017,Li2017} and have the advantage of not requiring additional qubit resources and can be implemented directly on noisy hardware. However, these techniques are only able to systematically remove the bias from the expectation values and cannot extend the coherence times of the computation. This manifests in a bad asymptotic scaling of most error mitigation methods emphasizing that such techniques are most favorable for near-term circuits.
In~\cite{suzuki2021quantum} it has been studied how to use error mitigation to correct for errors arising from approximate Solovay-Kitaev decompositions that reduce the number of $T$-gates. 

Given the significant hardware overhead required for the implementation of a fault-tolerant universal gate set it is meaningful to investigate the possibility of using error mitigation techniques to reduce this overhead in circuits using encoded qubits for the estimation of expectation values. In this work, we present two methods to mitigate the bias induced by noisy encoded $T$-gates using the \emph{quasi-probability decomposition} (QPD) method~\cite{Temme2017}. Both approaches circumvent the need for magic state distillation. The cost of the QPD method is an increased number of required shots to guarantee a fixed accuracy. For a single gate, this increase scales as $\gamma_{\eps}^2$ where $\gamma_{\eps}\geq 1$ is a quantity called \emph{sampling overhead} that captures the negativity of the quasi-probability decomposition. The sampling overhead depends on the error rate $\eps$ of the physical gates and on the error correcting code used for the encoding. It grows multiplicatively in the number of $T$-gates or $T$-count. For a circuit with $T$-count $t$, the total sampling overhead is $\Gamma:=\gamma_{\eps}^{t}$. In the limit $\eps \to 0$ the sampling overhead disappears, i.e., $\lim_{\eps \to 0} \gamma_{\eps} =1$. 

Our first method is based on noisy magic state preparation which has a favourable scaling of the sampling overhead but requires two logical qubits to simulate a logical $T$-gate. The second approach is based on code switching and requires a single logical qubit only, however at the cost of a larger sampling overhead. The two proposals achieve a sampling overhead that scales as $\gamma_{\eps} = 1 +2\kappa \eps + \cO(\eps^2)$, where $\kappa$ denotes a universal constant that varies for the two methods and depends on the considered error model as well as on the error correcting code. For realistic scenarios we find that $\kappa\approx 2/5$ and $\kappa \leq 30$ for the first and second method, respectively.

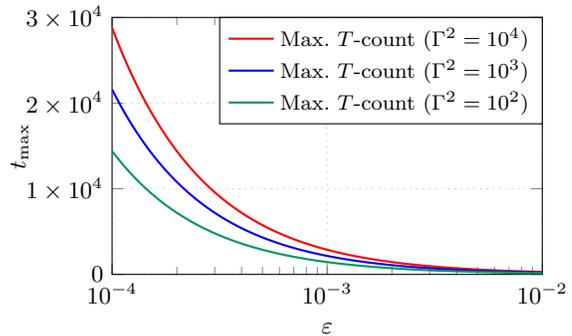
\begin{figure}[!htb]
 \begin{minipage}[t]{.5\textwidth}
    \centering
\begin{tikzpicture}
	\begin{axis}[
		height=5cm,
		width=7.3cm,
	    grid style=dotted,
		grid=major,
		xmode=log,
		xlabel=$\eps$,
		ylabel=$t_{\max}$,
		scaled y ticks = false,
		xmin=10^(-4),
		xmax=10^(-2),
		ymax=30000,
		ymin=0,
	    xtick={10^(-4),10^(-3),10^(-2)},
        ytick={30000,20000,10000,0},
         yticklabels={$3\times 10^4$,$2\times 10^4$,$1\times 10^4$, $0$},         
		legend style={at={(0.625,1.002)},anchor=north,legend cell align=left,font=\footnotesize} 
	]

	\addplot[red,thick,smooth] coordinates {
(0.0001,28780.) (0.00011,26163.4) (0.00012,23983.) (0.00013,22137.9) (0.00014,20556.5) (0.00015,19185.9) (0.00016,17986.6) (0.00017,16928.5) (0.00018,15987.9) (0.00019,15146.3) (0.0002,14388.9) (0.00021,13703.6) (0.00022,13080.6) (0.00023,12511.7) (0.00024,11990.3) (0.00025,11510.6) (0.00026,11067.8) (0.00027,10657.8) (0.00028,10277.1) (0.00029,9922.63) (0.0003,9591.8) (0.00031,9282.31) (0.00032,8992.17) (0.00033,8719.61) (0.00034,8463.08) (0.00035,8221.22) (0.00036,7992.78) (0.00037,7776.7) (0.00038,7571.99) (0.00039,7377.78) (0.0004,7193.28) (0.00041,7017.77) (0.00042,6850.63) (0.00043,6691.26) (0.00044,6539.13) (0.00045,6393.77) (0.00046,6254.72) (0.00047,6121.59) (0.00048,5994.01) (0.00049,5871.64) (0.0005,5754.16) (0.00051,5641.29) (0.00052,5532.76) (0.00053,5428.32) (0.00054,5327.76) (0.00055,5230.85) (0.00056,5137.4) (0.00057,5047.23) (0.00058,4960.16) (0.00059,4876.06) (0.0006,4794.75) (0.00061,4716.11) (0.00062,4640.01) (0.00063,4566.32) (0.00064,4494.93) (0.00065,4425.75) (0.00066,4358.65) (0.00067,4293.56) (0.00068,4230.39) (0.00069,4169.05) (0.0007,4109.46) (0.00071,4051.54) (0.00072,3995.24) (0.00073,3940.48) (0.00074,3887.2) (0.00075,3835.34) (0.00076,3784.84) (0.00077,3735.66) (0.00078,3687.74) (0.00079,3641.03) (0.0008,3595.49) (0.00081,3551.07) (0.00082,3507.74) (0.00083,3465.45) (0.00084,3424.16) (0.00085,3383.85) (0.00086,3344.48) (0.00087,3306.01) (0.00088,3268.41) (0.00089,3231.66) (0.0009,3195.73) (0.00091,3160.59) (0.00092,3126.21) (0.00093,3092.57) (0.00094,3059.65) (0.00095,3027.41) (0.00096,2995.85) (0.00097,2964.95) (0.00098,2934.67) (0.00099,2905.) (0.001,2875.93)
(0.0011,2614.27) (0.0012,2396.22) (0.0013,2211.72) (0.0014,2053.58) (0.0015,1916.52) (0.0016,1796.59) (0.0017,1690.77) (0.0018,1596.71) (0.0019,1512.55) (0.002,1436.81) (0.0021,1368.28) (0.0022,1305.98) (0.0023,1249.1) (0.0024,1196.96) (0.0025,1148.99) (0.0026,1104.71) (0.0027,1063.71) (0.0028,1025.64) (0.0029,990.189) (0.003,957.106) (0.0031,926.157) (0.0032,897.143) (0.0033,869.887) (0.0034,844.234) (0.0035,820.047) (0.0036,797.204) (0.0037,775.596) (0.0038,755.124) (0.0039,735.703) (0.004,717.253) (0.0041,699.703) (0.0042,682.988) (0.0043,667.051) (0.0044,651.838) (0.0045,637.302) (0.0046,623.397) (0.0047,610.084) (0.0048,597.326) (0.0049,585.089) (0.005,573.341) (0.0051,562.053) (0.0052,551.2) (0.0053,540.757) (0.0054,530.7) (0.0055,521.009) (0.0056,511.664) (0.0057,502.647) (0.0058,493.941) (0.0059,485.53) (0.006,477.399) (0.0061,469.535) (0.0062,461.924) (0.0063,454.556) (0.0064,447.417) (0.0065,440.498) (0.0066,433.789) (0.0067,427.28) (0.0068,420.963) (0.0069,414.828) (0.007,408.869) (0.0071,403.078) (0.0072,397.447) (0.0073,391.971) (0.0074,386.643) (0.0075,381.457) (0.0076,376.407) (0.0077,371.489) (0.0078,366.697) (0.0079,362.026) (0.008,357.471) (0.0081,353.03) (0.0082,348.696) (0.0083,344.467) (0.0084,340.339) (0.0085,336.308) (0.0086,332.37) (0.0087,328.523) (0.0088,324.764) (0.0089,321.089) (0.009,317.495) (0.0091,313.981) (0.0092,310.543) (0.0093,307.179) (0.0094,303.886) (0.0095,300.663) (0.0096,297.507) (0.0097,294.416) (0.0098,291.388) (0.0099,288.422) (0.01,285.514)
	};
	\addlegendentry{Max.~$T$-count ($\Gamma^2=10^4$)}	
	\addplot[thick,blue,smooth] coordinates {
(0.0001,21585.) (0.00011,19622.6) (0.00012,17987.2) (0.00013,16603.5) (0.00014,15417.4) (0.00015,14389.4) (0.00016,13490.) (0.00017,12696.4) (0.00018,11990.9) (0.00019,11359.7) (0.0002,10791.6) (0.00021,10277.7) (0.00022,9810.43) (0.00023,9383.81) (0.00024,8992.75) (0.00025,8632.97) (0.00026,8300.86) (0.00027,7993.36) (0.00028,7707.82) (0.00029,7441.97) (0.0003,7193.85) (0.00031,6961.74) (0.00032,6744.13) (0.00033,6539.71) (0.00034,6347.31) (0.00035,6165.91) (0.00036,5994.59) (0.00037,5832.53) (0.00038,5678.99) (0.00039,5533.33) (0.0004,5394.96) (0.00041,5263.33) (0.00042,5137.97) (0.00043,5018.44) (0.00044,4904.35) (0.00045,4795.33) (0.00046,4691.04) (0.00047,4591.2) (0.00048,4495.51) (0.00049,4403.73) (0.0005,4315.62) (0.00051,4230.97) (0.00052,4149.57) (0.00053,4071.24) (0.00054,3995.82) (0.00055,3923.13) (0.00056,3853.05) (0.00057,3785.42) (0.00058,3720.12) (0.00059,3657.04) (0.0006,3596.06) (0.00061,3537.08) (0.00062,3480.) (0.00063,3424.74) (0.00064,3371.2) (0.00065,3319.31) (0.00066,3268.99) (0.00067,3220.17) (0.00068,3172.79) (0.00069,3126.79) (0.0007,3082.09) (0.00071,3038.66) (0.00072,2996.43) (0.00073,2955.36) (0.00074,2915.4) (0.00075,2876.5) (0.00076,2838.63) (0.00077,2801.74) (0.00078,2765.8) (0.00079,2730.77) (0.0008,2696.61) (0.00081,2663.3) (0.00082,2630.8) (0.00083,2599.08) (0.00084,2568.12) (0.00085,2537.89) (0.00086,2508.36) (0.00087,2479.51) (0.00088,2451.31) (0.00089,2423.75) (0.0009,2396.8) (0.00091,2370.44) (0.00092,2344.66) (0.00093,2319.43) (0.00094,2294.73) (0.00095,2270.56) (0.00096,2246.89) (0.00097,2223.71) (0.00098,2201.) (0.00099,2178.75) (0.001,2156.95)
(0.0011,1960.7) (0.0012,1797.17) (0.0013,1658.79) (0.0014,1540.18) (0.0015,1437.39) (0.0016,1347.44) (0.0017,1268.08) (0.0018,1197.54) (0.0019,1134.42) (0.002,1077.61) (0.0021,1026.21) (0.0022,979.487) (0.0023,936.826) (0.0024,897.719) (0.0025,861.741) (0.0026,828.531) (0.0027,797.781) (0.0028,769.227) (0.0029,742.642) (0.003,717.83) (0.0031,694.618) (0.0032,672.857) (0.0033,652.415) (0.0034,633.175) (0.0035,615.035) (0.0036,597.903) (0.0037,581.697) (0.0038,566.343) (0.0039,551.777) (0.004,537.94) (0.0041,524.777) (0.0042,512.241) (0.0043,500.288) (0.0044,488.879) (0.0045,477.976) (0.0046,467.548) (0.0047,457.563) (0.0048,447.994) (0.0049,438.816) (0.005,430.005) (0.0051,421.54) (0.0052,413.4) (0.0053,405.568) (0.0054,398.025) (0.0055,390.757) (0.0056,383.748) (0.0057,376.985) (0.0058,370.455) (0.0059,364.147) (0.006,358.049) (0.0061,352.151) (0.0062,346.443) (0.0063,340.917) (0.0064,335.563) (0.0065,330.374) (0.0066,325.342) (0.0067,320.46) (0.0068,315.722) (0.0069,311.121) (0.007,306.652) (0.0071,302.308) (0.0072,298.085) (0.0073,293.978) (0.0074,289.982) (0.0075,286.093) (0.0076,282.306) (0.0077,278.617) (0.0078,275.022) (0.0079,271.519) (0.008,268.104) (0.0081,264.772) (0.0082,261.522) (0.0083,258.35) (0.0084,255.254) (0.0085,252.231) (0.0086,249.278) (0.0087,246.392) (0.0088,243.573) (0.0089,240.817) (0.009,238.122) (0.0091,235.486) (0.0092,232.907) (0.0093,230.384) (0.0094,227.915) (0.0095,225.497) (0.0096,223.13) (0.0097,220.812) (0.0098,218.541) (0.0099,216.316) (0.01,214.136)
	};
	\addlegendentry{Max.~$T$-count ($\Gamma^2=10^3$)}	
	
	\addplot[ForestGreen,thick,smooth] coordinates {
(0.0001,14390.) (0.00011,13081.7) (0.00012,11991.5) (0.00013,11069.) (0.00014,10278.2) (0.00015,9592.95) (0.00016,8993.32) (0.00017,8464.24) (0.00018,7993.94) (0.00019,7573.14) (0.0002,7194.43) (0.00021,6851.78) (0.00022,6540.28) (0.00023,6255.87) (0.00024,5995.16) (0.00025,5755.31) (0.00026,5533.91) (0.00027,5328.91) (0.00028,5138.55) (0.00029,4961.32) (0.0003,4795.9) (0.00031,4641.16) (0.00032,4496.09) (0.00033,4359.81) (0.00034,4231.54) (0.00035,4110.61) (0.00036,3996.39) (0.00037,3888.35) (0.00038,3786.) (0.00039,3688.89) (0.0004,3596.64) (0.00041,3508.89) (0.00042,3425.31) (0.00043,3345.63) (0.00044,3269.57) (0.00045,3196.88) (0.00046,3127.36) (0.00047,3060.8) (0.00048,2997.01) (0.00049,2935.82) (0.0005,2877.08) (0.00051,2820.64) (0.00052,2766.38) (0.00053,2714.16) (0.00054,2663.88) (0.00055,2615.42) (0.00056,2568.7) (0.00057,2523.61) (0.00058,2480.08) (0.00059,2438.03) (0.0006,2397.37) (0.00061,2358.05) (0.00062,2320.) (0.00063,2283.16) (0.00064,2247.47) (0.00065,2212.87) (0.00066,2179.33) (0.00067,2146.78) (0.00068,2115.2) (0.00069,2084.52) (0.0007,2054.73) (0.00071,2025.77) (0.00072,1997.62) (0.00073,1970.24) (0.00074,1943.6) (0.00075,1917.67) (0.00076,1892.42) (0.00077,1867.83) (0.00078,1843.87) (0.00079,1820.51) (0.0008,1797.74) (0.00081,1775.53) (0.00082,1753.87) (0.00083,1732.72) (0.00084,1712.08) (0.00085,1691.93) (0.00086,1672.24) (0.00087,1653.) (0.00088,1634.21) (0.00089,1615.83) (0.0009,1597.87) (0.00091,1580.29) (0.00092,1563.1) (0.00093,1546.28) (0.00094,1529.82) (0.00095,1513.71) (0.00096,1497.93) (0.00097,1482.47) (0.00098,1467.33) (0.00099,1452.5) (0.001,1437.96)
(0.0011,1307.14) (0.0012,1198.11) (0.0013,1105.86) (0.0014,1026.79) (0.0015,958.259) (0.0016,898.296) (0.0017,845.387) (0.0018,798.357) (0.0019,756.277) (0.002,718.406) (0.0021,684.141) (0.0022,652.992) (0.0023,624.55) (0.0024,598.48) (0.0025,574.494) (0.0026,552.354) (0.0027,531.854) (0.0028,512.818) (0.0029,495.095) (0.003,478.553) (0.0031,463.079) (0.0032,448.571) (0.0033,434.943) (0.0034,422.117) (0.0035,410.024) (0.0036,398.602) (0.0037,387.798) (0.0038,377.562) (0.0039,367.852) (0.004,358.626) (0.0041,349.851) (0.0042,341.494) (0.0043,333.525) (0.0044,325.919) (0.0045,318.651) (0.0046,311.699) (0.0047,305.042) (0.0048,298.663) (0.0049,292.544) (0.005,286.67) (0.0051,281.027) (0.0052,275.6) (0.0053,270.378) (0.0054,265.35) (0.0055,260.504) (0.0056,255.832) (0.0057,251.323) (0.0058,246.97) (0.0059,242.765) (0.006,238.699) (0.0061,234.767) (0.0062,230.962) (0.0063,227.278) (0.0064,223.709) (0.0065,220.249) (0.0066,216.895) (0.0067,213.64) (0.0068,210.481) (0.0069,207.414) (0.007,204.435) (0.0071,201.539) (0.0072,198.724) (0.0073,195.986) (0.0074,193.322) (0.0075,190.728) (0.0076,188.204) (0.0077,185.744) (0.0078,183.348) (0.0079,181.013) (0.008,178.736) (0.0081,176.515) (0.0082,174.348) (0.0083,172.234) (0.0084,170.169) (0.0085,168.154) (0.0086,166.185) (0.0087,164.262) (0.0088,162.382) (0.0089,160.544) (0.009,158.748) (0.0091,156.99) (0.0092,155.271) (0.0093,153.589) (0.0094,151.943) (0.0095,150.332) (0.0096,148.754) (0.0097,147.208) (0.0098,145.694) (0.0099,144.211) (0.01,142.757)
	};
	\addlegendentry{Max.~$T$-count ($\Gamma^2=10^2$)}

	\end{axis}  
\end{tikzpicture}
\end{minipage}

    \caption{Maximal $T$-count that can be performed via the magic state method with a total sampling overhead $\Gamma^2 \in\{10^2,10^3,10^4\}$. We assume a depolarizing noise model with only two-qubit errors, i.e., $\gamma_{\eps}$ is given in~\eqref{eq_gamma_approx} for $\kappa=2/5$, where we neglect second order error terms.}
    \label{fig_TcountSamplingOverhead}
\end{figure}

As a result, assuming that we fix some value of the total sampling overhead $\Gamma^2$ that the user is willing to accept, we can simulate the outcome of a universal fault-tolerant quantum circuit with a $T$-count up to $\eta/\eps+\cO(1)$, where $\eta:=\log(\Gamma^2)/(4\kappa)$.
More concretely, for a physical error rate $\eps = 10^{-2}$ and total sampling overhead $\Gamma^2=10^3$ we can simulate outcomes for circuits with up to 200 $T$-gates. In case of further experimental progress such that $\eps = 10^{-3}$ we can go up to 2000 $T$-gates. We refer to Figure~\ref{fig_TcountSamplingOverhead} for more details and the precise calculations. This considerably exceeds the possibilities of state of the art classical algorithms to simulate universal quantum circuits~\cite{BG16,pashayan21} which can deal with approximately 50 $T$-gates --- assuming that the number of qubits is sufficiently large such that brute-force calculations are not feasible.

\textit{Error-mitigated logical T-gates via noisy magic states.}---
One possibility to implement a $T$-gate, i.e., $T=\diag(1,\ee^{\ci \pi/4})$, is via a Clifford circuit shown in Figure~\ref{fig_Tgate} and a magic state $\ket{\pi/4} = ( \ket{0} + \ee^{\ci \pi/4} \ket{1})/\sqrt{2}$.  
This approach reduces the problem of implementing a fault-tolerant $T$-gate to the task of a fault-tolerant preparation of an encoded magic state. The latter task can be achieved by magic state distillation~\cite{BK05}, where several noisy magic states are transformed into fewer magic states of better fidelity. Whereas magic state distillation is a very elegant approach for achieving universal fault-tolerant quantum computing, the distillation process leads to a considerable overhead in practice. In our method we forego the usage of magic state distillation and make use of the same circuit in Figure~\ref{fig_Tgate} with a noisy magic state. This will lead to a faulty $T$-gate. In a second step, we use quantum error mitigation techniques, such as the QPD method, to suppress this error.

Let $\rho$ be a noisy encoded magic state. We define the logical error rate as
\begin{align} \label{eq_epsBar}
\bar\eps:=1-\bra{\pi/4}\rho \ket{\pi/4}   \, . 
\end{align}
For simplicity, assume that the chosen error correction code enables a noiseless implementation of logical gates $S = \diag(1,\ci)$, $X$, $\CNOT$, preparation of logical $\ket{0}$ and $\ket{+}$ states, and measurements of a logical qubit in the $Z$ and $X$ basis~\footnote{This assumption can be forced to hold by choosing a code that has a sufficiently large distance. Note that these gates are Clifford gates that feature a transversal and hence fault-tolerant implementation for self-dual CSS codes.}.
We can twirl the noisy magic state $\rho$ with respect to the group $\{I,A\}$, where $A$ is the Clifford gate defined as
\begin{align*}
    A:=e^{-\ci\pi/4}SX = \proj{\pi/4}-\proj{\omega} \, ,
\end{align*}
where $\ket{\omega}:=Z\ket{\pi/4}$ is a state orthogonal to $\ket{\pi/4}$ and $I$ denotes the identity. The twirled state thus becomes
\begin{align} \label{eq_twirledState}
    \tau:=\frac{1}{2}(\rho + A\rho A^\dagger) = (1-\bar\eps)\proj{\pi/4} + \bar\eps \proj{\omega} \, ,
\end{align}
where the second equality follows from the identity $A\ket{\pi/4}\bra{\omega}A^{\dagger} + \ket{\pi/4}\bra{\omega}=0$. 
Thus we can prepare the twirled state $\tau$ by first preparing $\rho$ and then applying the logical $A$ gate with probability $1/2$. The twirled state $\tau$ can be considered as the ideal magic state $\ket{\pi/4}$ that suffers from a random Pauli $Z$ error applied with probability $\bar \eps$, since $\ket{\omega}= Z\ket{\pi/4}$.

Note that a Pauli $Z$ error on the magic state $\ket{\pi/4}$ has no effect on the measurement in Figure~\ref{fig_Tgate}. Thus using the twirled state $\tau$ instead of the ideal magic state $\ket{\pi/4}$ inside the $T$-gate gadget in Figure~\ref{fig_Tgate}, one can implement a noisy $T$-gate
\begin{align}\label{eq_noisyT}
    \cT_{\bar \eps} = (1-\bar \eps) \cT + \bar \eps \, \cZ \circ \cT \, ,
\end{align}
where $\cT$ and $\cZ$ denote the quantum channels implementing the ideal $T$- and $Z$-gate, respectively, i.e., $\cT(B)=TBT^\dagger$ and $\cZ(B)=ZBZ^\dagger$ for any single qubit operator $B$.
To use error mitigation to simulate a perfect $T$-gate using $\cT_{\bar \eps}$, we need a good estimate for the logical error rate $\bar \eps$. This could be done via standard tomography techniques, however we present a more efficient method below.

It remains to understand how the logical error rate $\bar \eps$ defined in~\eqref{eq_epsBar} depends on the physical error rate $\eps$. It has been shown~\cite{horodecki15} that there exist noisy state preparation circuits for the surface code that prepare a noisy encoded magic state $\rho$ such that the logical error rate is upper bounded by a constant that is independent of the code distance~\cite{horodecki15}. In fact $\bar\eps$ can be in the same order as the physical error rate $\eps$. For the surface code and a depolarizing error model it has been shown~\cite{Li_2015} that
\begin{align} \label{eq_logical_ER}
    \bar \eps = \kappa \eps + \cO(\eps^2) \, ,
\end{align}
where $\kappa$ is a conversion constant that depends on the details of the noise model. For example, in case of perfect single qubit gates and no initialization errors we find $\kappa=2/5$~\cite{Li_2015}.
Furthermore, the second order term in~\eqref{eq_logical_ER} is negligible for realistic error rates.\footnote{The protocol in~\cite{Li_2015} uses post-selection to suppress logical error with a relatively large success rate. Hence the cost of this step is modest. More information can be found in~\cite{Li_2015}.}

The standard Kitaev's surface code does not feature a transversal logical $S$-gate because it is not a self-dual CSS code. In fact, it is known that the surface code does not permit logical non-Pauli gates realizable
by geometrically local constant-depth circuits~\cite[Theorem~7.1]{beverland2016protected}.
As a result, it is not clear how to implement the logical Clifford operator $A$ required for the twirling~\eqref{eq_twirledState}. Luckily any logical 
single-qubit Clifford gate  
for the surface code
can be implemented 
by a constant depth 
circuit with long-range gates~\cite{moussa16,BGKT20}. 
Any constant-depth circuit
is automatically fault-tolerant
even if it is not geometrically local~\cite{BGKT20}.
Indeed, the constant depth condition 
implies that 
a single fault in the circuit can affect only a constant number of output qubits. 
This gives a fault-tolerant implementation of the $S$-gate for the Kitaev's surface code.

To experimentally determine the logical error rate $\bar\eps$ we consider a  phase flip channel $\cN_{\bar \eps}:= (1-\bar \eps) \cI + \bar \eps \cZ$, where $\cI$ denotes the identity channel. Since $\cT$ and $\cZ$ commute we have for any $p \in \N$
\begin{equation} \label{eq_commute}
    \cT_{\bar \eps}^p = (\cN_{\bar \eps} \circ \cT)^p = \cN_{\bar \eps}^p \circ \cT^p\, .
\end{equation}
For $p\equiv 0 \,\, (\!\!\!\!\mod 8)$ we have $T^p=\mathds{1}$ and $\cT^p=\cI$. Hence~\eqref{eq_commute} ensures that $\cT_{\bar \eps}^p=\cN_{\bar \eps}^p$. We can now learn $\bar \eps$ by preparing a logical state $\ket{+}$ fault-tolerantly, applying $\cT^p_{\bar \eps}$ and measuring the output in the $\{\ket{+},\ket{-}\}$ basis, which gives us an outcome $1$ or $0$. Hence we can estimate the expectation value of the outcome 
\begin{align} \label{eq_tomography}
    f(p)
    := \bra{+} \cT^p_{\bar \eps}(\proj{+}) \ket{+}
    =\frac{1}{2}\Big(1+(1-2\bar\eps)^{p}\Big) \, ,
\end{align}
with an approximation error that decreases exponentially in the number of circuit shots~\footnote{The final step in~\eqref{eq_tomography} can be checked with a simple iterative argument.}.
By measuring $f(p)$ for $p=8k$ with $k \in \N$, we can obtain a good estimate for $\bar \eps$ using exponential fitting.

Once we learned the logical error rate $\bar \eps$ we can use various error mitigation techniques, such as the QPD method~\cite{Temme2017,Endo2018,PSW21}, to transform the noisy $T$-gate $\cT_{\bar \eps}$ into a good approximation of the perfect $T$-gate $\cT$. 
The QPD method allows for the simulation of an arbitrary channel $\mathcal{F}$ while only having access to quantum hardware that can execute the noisy quantum channels $\{\mathcal{E}_i\}$. The central ingredient of the method is a decomposition $\mathcal{F} = \sum_i a_i \mathcal{E}_i$, where the $a_i\in\mathbb{R}$ are quasi-probability coefficients. During execution of the circuit, the gate is probabilistically replaced with one of the channels $\mathcal{E}_i$ with probability $\abs{a_i} / \gamma$ where $\gamma:=\sum_i \abs{a_i}$. By correctly weighting the measurement outcome at the end of the circuit, one can obtain an unbiased estimate of the true expectation value of the outcome of the ideal quantum circuit by performing Monte Carlo sampling. The number of samples required to reach a certain accuracy $\delta>0$ scales as $\cO(\gamma^2/\delta^2)$. We therefore call $\gamma$ the \emph{sampling overhead}. 

We can express the perfect $T$-gate $\cT$ in terms of the noisy $T$-gate $\cT_{\bar \eps}$ by using the identity
\begin{align*}
    \cT = \left( \frac{1-\bar \eps}{1-2\bar \eps} \right) \cT_{\bar \eps} - \left( \frac{\bar \eps}{1-2\bar \eps} \right) \cZ \circ \cT_{\bar \eps} \, ,
\end{align*}
which is a QPD with sampling overhead $\gamma= 1/(1-2\bar \eps)$. Expanding this term around $\bar \eps = 0$ gives 
\begin{align} \label{eq_gamma_approx}
    \gamma_{\eps}= 1+2\bar \eps + \cO(\bar \eps^2) = 1 + 2\kappa \eps + \cO(\eps^2) \, , 
\end{align}
if $\bar\eps \ll 1$ and $\eps \ll 1$, where $\kappa$ denotes a conversion constant that depends on the used error-correcting code as well as on the error model. The final step in~\eqref{eq_gamma_approx} uses the fact that it is possible to do encoding such that the logical error rate $\bar \eps$ is proportional to the physical error rate $\eps$ up to quadratic terms, in the sense of~\eqref{eq_logical_ER}. 
To keep our framework general we treat $\kappa$ an arbitrary constant so that various codes or underlying physical error models fit our setting. 

\textit{Error-mitigated logical T-gates via code switching.}---
The implementation of a $T$-gate via magic states (see Figure~\ref{fig_Tgate}) has the drawback of using two encoded qubits. 
 It also includes a logical CNOT gate which may require more logical qubits
for certain  implementations~\cite{horsman2012surface}. 
One may ask if a logical $T$-gate can be implemented using only one
logical qubit. 
In this section we show how to accomplish this task by combining
the QPD method and the code switching method pioneered by Paetznick
and Reichardt~\cite{paetznick2013universal}.

Suppose $\calS_1$ and $\calS_2$ are CSS-type~\cite{calderbank1996good,steane1996multiple} 
quantum codes with  one logical qubit.
Our goal is to implement a logical $T$-gate on a qubit encoded by $\calS_1$.
We assume that $\calS_1$ has a large distance for both $X$ and $Z$ errors and
enables a fault-tolerant implementation of  the Clifford gate $S=T^2$.
We assume that $\calS_2$ is an asymmetric  code
with a large distance for $X$ errors and a distance $1$ for $Z$ errors
such that a logical-$Z$ operator of $\calS_2$ can be chosen as a single-qubit
Pauli $Z$ on some physical qubit. We shall denote this local logical-$Z$ operator
as  $\overline{Z}_{\mathrm{loc}}$. 
This allows us to perform a logical $T$-gate 
on a qubit encoded by $\calS_2$ simply by applying a physical $T$-gate on the
qubit acted upon by $\overline{Z}_{\mathrm{loc}}$. Indeed, since the $T$-gate is a linear
combination of the identity and the Pauli $Z$, the physical and the logical $T$-gates
become equivalent. The code switching method enables a conversion between
the codes $\calS_1$ and $\calS_2$ by measuring stabilizers of $\calS_2$ on a logical
state encoded by $\calS_1$ or vice versa. In certain cases this conversion can
be performed fault-tolerantly~\cite{paetznick2013universal}.
This is achieved by identifying stabilizers
present in both codes and using the measured syndromes of such stabilizers
to diagnose and correct errors. 
The error suppression  typically scales exponentially with  the code distance.
In our case the code switching protects the encoded qubit 
exponentially well only from logical $X$ errors. This ensures that the effective noise
channel acting on the logical qubit is dominated by $Z$-type (possibly coherent) errors.
Thus the code switching implements a noisy logical $T$-gate
of the form $\mathcal{T}_n=\mathcal{N}\circ\mathcal{T}$ where $\mathcal{N}$ is some
$Z$-type noise channel. We can tailor this noise to become stochastic $Z$ noise by twirling over the $X$ gate using the identity 
\begin{align*}
  \frac{1}{2}(\mathcal{N} + \mathcal{X}\circ\mathcal{N}\circ\mathcal{X}) = (1-\bar\eps)\cI + \bar\eps \mathcal{Z} =:  \mathcal{N}_{\bar \eps} \, ,
\end{align*}
for some logical error rate $\bar \eps$ where
$\cX(B)=XBX^\dagger$ for any single qubit operator $B$.
Using the identity $XT=TA$ we get
\begin{align}\label{eq_deformationTwirling}
    \frac{1}{2}(\mathcal{T}_n + \cX\circ\mathcal{T}_n\circ\cA) =
    \mathcal{N}_{\bar \eps}\circ\mathcal{T} =\cT_{\bar \eps} \, ,
\end{align}
where $\cA(B)=ABA^\dagger$ for any single qubit operator $B$ and $\cT_{\bar \eps}$ is defined as in~\eqref{eq_noisyT}.
By assumption, the code $\calS_1$ enables a fault-tolerant implementation
of the Pauli $X$ and the $S$ gate on the encoded qubit. Since $A=SX$ up to an overall phase,
the twirling in~\eqref{eq_deformationTwirling} can be implemented fault-tolerantly
on a qubit encoded by $\calS_1$
by applying either $\mathcal{T}_n$ or $\cX\circ\mathcal{T}_n\circ\cA$ with the probability
$1/2$ each. 
Since at this point we 
 have implemented a noisy $T$-gate of the form $\cT_{\bar \eps}$, we can
follow the procedure discussed above to determine the parameter $\bar \eps$ efficiently. The error mitigation procedure is completely analogous as before.

In Appendix~\ref{app_kitaev} we specialize the code switching method to
Kitaev's surface code~\cite{kitaev06} depicted on Figure~\ref{fig:Kitaev}.
It is shown that the logical error rate scales as $\bar \eps \approx 30 \eps$ for realistic error models. 

\begin{figure}
\centerline{
\includegraphics[width=0.45\textwidth]{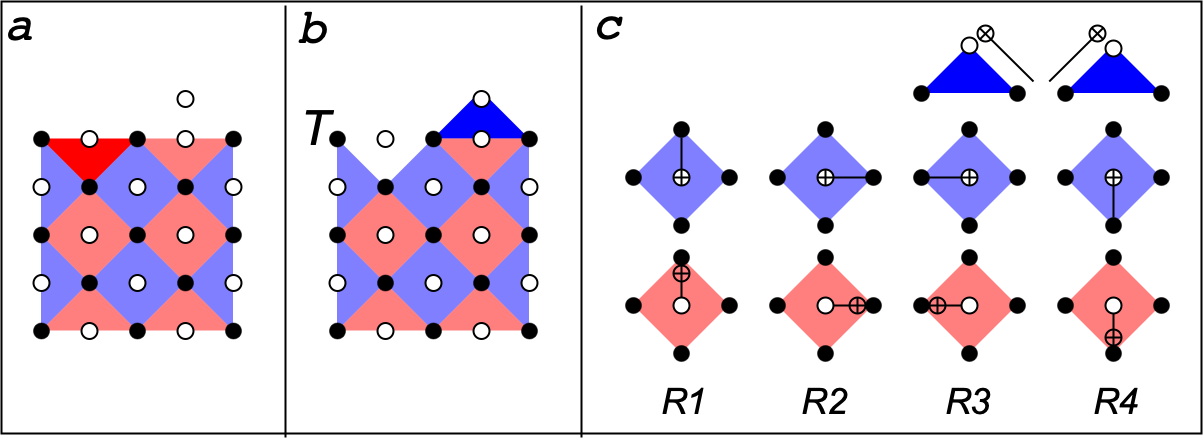}}
\caption{Logical $T$-gate by code switching.
(a) Surface code $\calS_1$ with the distance $d=3$.
Black and white circles indicate data and syndrome
qubits. Red and blue faces indicate $X$ and $Z$ stabilizers.
(b) Asymmetric surface code $\calS_2$ exhibiting a single-qubit
logical $T$-gate at the north-west corner of the lattice.
Code switching $\calS_1\to \calS_2$ is performed by 
turning off $X$ stabilizer $F_1$ (dark red triangle)
and turning on $Z$ stabilizer $G_1$ (dark blue triangle).
(c) The syndrome extraction cycle consists of four rounds of CNOTs R1, R2, R3, R4. To avoid clutter we only show a local schedule of CNOTs for each stabilizer.
Schedules for weight-3 stabilizers are properly
truncated. 
The schedule extends to the full lattice
in a translation invariant fashion.
\label{fig:Kitaev}
}
\end{figure}

\textit{Comparison with QEC and classical simulation.}---
Using the schemes explained above we can simulate universal fault-tolerant quantum circuits with a total sampling overhead $\Gamma^2 = \gamma_{\eps}^{2t}$, where $t$ denotes the $T$-count of the circuit. Recalling~\eqref{eq_gamma_approx} and setting the total sampling overhead $\Gamma^2$ to a fixed desired value gives for $\eta:=\log(\Gamma^2)/(4\kappa)$
\begin{align*}
    t 
    = \frac{\log \Gamma^2}{2 \log(1+2\kappa \eps +\cO(\eps^2))} 
    =  \eta/\eps + \cO(1) \quad \textnormal{for} \quad \eps \ll 1 \, .
\end{align*}
If we neglect the second order term in~\eqref{eq_gamma_approx} we find $t\approx \eta(1/\eps + \kappa)$, which is plotted in Figure~\ref{fig_TcountSamplingOverhead}. 
We see that for realistic scenarios our scheme allows the simulation of quantum circuits with a number of $T$-gates that are unfeasible for classical algorithms~\cite{BG16}.

Table~\ref{tab_overview} compares the presented approach to conventional quantum error correction (QEC) and classical simulation. The new method comes with a simulation overhead that scales exponentially in the number of $T$-gates however the base of this exponential function depends on the physical error rate $\eps$ and is considerably smaller compared to classical simulation for realistic noise levels. 
\begin{table}[!htb]
    \centering
    \begin{tabular}{c | c c c} \hline
                            &  QEC & QPD \& QEC & classical sim.    \\ \hline \\[-0.3cm]
      \multirow{2}{*}{hardware} & fault-tolerant Clif-  &fault-tolerant  &classical\\
       &ford \& magic state & Clifford & computer \\ \\[-0.3cm]
  $T$-gate\\ overhead & \multirow{-2}{*}{$\mathrm{const}$} &  \multirow{-2}{*}{$\gamma_{\eps}^{2t}$} & \multirow{-2}{*}{$1.3831^t$} \\\hline
    \end{tabular}
    \caption{Comparison of the new approach with conventional quantum error correction and classical simulation.}
    \label{tab_overview}
\end{table}

It is an open question for future research to figure out how error mitigation techniques can be efficiently used for non-transversal gates beyond the $T$-gate. The two methods proposed in this paper cannot be directly generalized to arbitrary non-transversal gates, as the procedure of twirling the noise relies on the fact that the $T$-gate lies in the third Clifford hierarchy. However, if one were to forgo the tailoring of the noise to a Pauli channel, the code switching method could be used to implement an arbitrary noisy $R_z$ rotation. This comes with the drawback that the characterization of the (possibly coherent) noise would presumably require a full tomography procedure, which is significantly less efficient than the exponential fitting method described above.

In Appendix~\ref{app_limitations}, we discuss how to deal with errors in the fault-tolerant Clifford gates which can be exponentially suppressed by increasing the blocklength of the error correcting code used.


\emph{Note.---} During the preparation of this work, we became aware of an independent effort to use error mitigation for universal quantum computing via encoded Clifford+$T$ circuits~\cite{CL21}. 

\emph{Acknowledgements.---} We thank Stefan Woerner for helpful comments and discussions. Part of this work was done when CP~was a research intern at IBM Research Zurich. CP acknowledges support from the Swiss National Science Foundation via the National Centre of Competence in Research QSIT. SB is supported in part by the IBM Research Frontiers Institute.

\onecolumngrid
\appendix

\section{Code switching construction for Kitaev's surface code} \label{app_kitaev}
Above we described how to realize error-mitigated logical T-gates via code switching. In this appendix we present a detailed construction for Kitaev's surface code~\cite{kitaev06} depicted on Figure~\ref{fig:Kitaev}.
A distance-$d$ code surface code is defined on a square lattice of linear size $2d-1$.
Code qubits live at sites indicated by black circles
on Figure~\ref{fig:Kitaev}.
Let $\calS_1$ be the stabilizer group of the surface code generated by $X$ and $Z$ stabilizers
located on red ($X$) and blue ($Z$) faces of the lattice,
see Figure~\ref{fig:Kitaev}.
We choose logical Pauli operators $\overline{X}$ and $\overline{Z}$ as products of single-qubit $X$ and $Z$
along  the left ($X$)
and the top ($Z$) boundary.

\begin{figure}[!htb]
\centerline{
\includegraphics[width=0.9\textwidth]{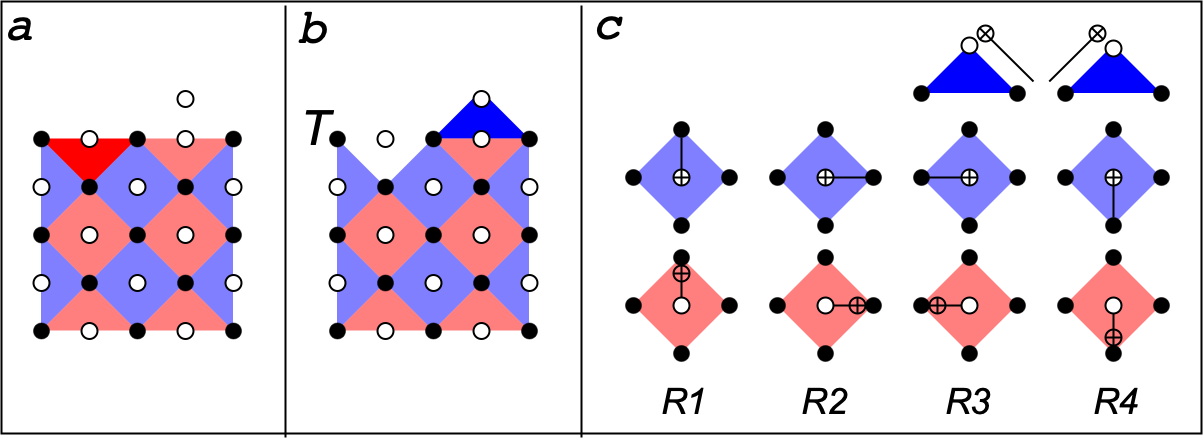}}
\caption{\textbf{Logical $T$-gate by code switching.}
(a) Surface code $\calS_1$ with the distance $d=3$.
Black and white circles indicate data and syndrome
qubits. Red and blue faces indicate $X$ and $Z$ stabilizers.
(b) Asymmetric surface code $\calS_2$ exhibiting a single-qubit
logical $T$-gate at the north-west corner of the lattice.
Code switching $\calS_1\to \calS_2$ is performed by 
turning off $X$ stabilizer $F_1$ (dark red triangle)
and turning on $Z$ stabilizer $G_1$ (dark blue triangle).
(c) The syndrome extraction cycle consists of four rounds of CNOTs R1, R2, R3, R4. To avoid clutter we only show a local schedule of CNOTs for each stabilizer.
Schedules for weight-3 stabilizers are properly
truncated. 
The schedule extends to the full lattice
in a translation invariant fashion.
\label{fig:Kitaev}
}
\end{figure}

Next let us define an asymmetric version of the surface code
with a small distance for $Z$ errors.
Its stabilizer group $\calS_2$ is obtained from $\calS_1$ by introducing  some new $Z$ stabilizers
and removing some $X$  stabilizers. This modification only affects
stabilizers located near the top boundary,
as shown on Figure~\ref{fig:Kitaev}.
To describe this formally, let $d=2t+1$
be the code distance. 
Label code qubits located at the
top boundary by integers $1,2,\ldots,d$
in the order from the left to the right such that $\overline{Z}=Z_1Z_2\cdots Z_d$.
Define Pauli operators 
\begin{align*}
 G_i := Z_{2i} Z_{2i+1}.
\end{align*}
The group $\calS_2$ is obtained from $\calS_1$ by 
adding $t$ new stabilizers $G_1,\ldots,G_t$ and
removing every second $X$ stabilizer adjacent to the top boundary,
see Figure~\ref{fig:Kitaev} for an example. 
An $X$ stabilizer of $\calS_1$ is removed if it
anti-commutes with some operator $G_i$.
Let $q_{\mathrm{loc}}:= 1$ be the data qubit located at the
north-west corner of the lattice.
The code $\calS_2$ exhibits a local logical-$Z$ operator
acting only on the qubit $q_{\mathrm{loc}}$, namely
\begin{equation}
\label{deform0}
\overline{Z}_{\mathrm{loc}} =
\overline{Z}\prod_{i=1}^t  G_i = Z_{q_{\mathrm{loc}}}.
\end{equation}
Note that $\overline{Z}_{\mathrm{loc}}$ and $\overline{Z}$ differ by 
stabilizers of $\calS_2$ and thus have the same action on any logical state
of $\calS_2$.
Thus the logical $T$-gate on a qubit encoded by $\calS_2$ can be implemented
by either of the following operators:
\begin{equation}
\label{deform1}
\overline{T} = \ee^{-\ci \frac{\pi}{8} \overline{Z}}
\quad \mbox{and}
\quad \overline{T}_{\mathrm{loc}} = \ee^{-\ci \frac{\pi}{8} \overline{Z}_{\mathrm{loc}}} \, .
\end{equation}
Here we ignore the overall phase.
Note that $\overline{T}$ also implements a logical $T$-gate for the code
$\calS_1$ since both codes have the same logical Pauli operators.
Let $\calG=\la G_1,\ldots,G_t\ra$ be the group generated by 
$G_1,\ldots,G_t$.
We claim that a logical $T$-gate on a qubit
encoded by the surface code $\calS_1$ can be implemented
as follows. 

\begin{algorithm}[H]
	\caption{Logical $T$-gate by code switching}
	\label{algo_1}
	\begin{algorithmic}[1]
    \State{Initialize code qubits in a logical state of the surface code $\calS_1$}
    \State{Measure the syndrome of the asymmetric surface code $\calS_2$}
    \State{Let $\sigma_i \in \{-1,1 \}$ be the measured syndrome of $G_i$}
    \State{Compute $\sigma=\prod_{i=1}^t \sigma_i$}
    \State{Apply a single-qubit operator $(\overline{T}_{\mathrm{loc}})^\sigma$
    to the post-measurement state}
    \State{Measure the syndrome of the surface code $\calS_1$}
    \State{Compute a Pauli operator $R\in \calG$ consistent with the measured syndrome of $\calS_1$}
    \State{Apply $R$ to the post-measurement state}
    \end{algorithmic}
\end{algorithm}
Indeed, let $|\psi_j\ra$ be a state obtained after executing 
the first $j$ steps of this algorithm. 
We need to show that $|\psi_8\ra = \overline{T} |\psi_1\ra$.
Let $\Pi$ be a projector describing the measurement at step~2
such that $|\psi_2\ra=\Pi |\psi_1\ra$, ignoring the normalization.
From $G_i|\psi_2\ra=\sigma_i|\psi_2\ra$
and~\eqref{deform0},\eqref{deform1} one infers 
$(\overline{T}_{\mathrm{loc}})^\sigma |\psi_2\ra=
\overline{T} |\psi_2\ra$.
Since $\overline{Z}$ is a logical operator for both codes $\calS_i$,
the logical gate $\overline{T}$
commutes with the syndrome measurement of $\calS_2$.
Thus 
\begin{align*}
|\psi_5\ra=
(\overline{T}_{\mathrm{loc}})^\sigma |\psi_2\ra =
\overline{T} |\psi_2\ra=\Pi \overline{T} |\psi_1\ra \, .
\end{align*}
Let $\Lambda$ be a projector describing the measurement at step~6. In the absence of errors
the measured syndrome is non-trivial only for
stabilizers of $\calS_1$ that anti-commute with some
element of $\calS_2$.
One can check that there are exactly $t$ such stabilizers
which we denote $F_1,\ldots,F_t$.
The stabilizer $F_i$ acts on data qubits
$2i-1,2i$ located at the top boundary as well
as the data qubit located at the second row
between $2i-1$ and $2i$, see Figure~\ref{fig:Kitaev}.
Let $\lambda_i\in \{1,-1\}$ be the syndrome of $F_i$
measured at step~6. The operator $R$ used at step~7 is defined as
\begin{align*}
R=\prod_{i\, : \, \lambda_i=-1} 
\; \prod_{a=i}^t G_a \, . 
\end{align*}
Using the commutation rules between the operators $F_i$ and $G_a$
one can easily check that $R$ has the syndrome $\lambda$,
that is, $RF_i=\lambda_i F_i R$ for all $i=1,\ldots,t$.
Write
$\Lambda = R \Gamma R$, where $\Gamma$ is a projector onto the logical subspace of the surface code $\calS_1$.
The above shows that 
\begin{align*}
|\psi_8\ra = R\Lambda |\psi_5\ra=
(R\Lambda R)R |\psi_5\ra=\Gamma R |\psi_5\ra=
\Gamma R \Pi \overline{T} |\psi_1\ra=
\Gamma (R\Pi) \Gamma \overline{T} |\psi_1\ra \, .
\end{align*}
To obtain the last equality we
noted that $\Gamma |\psi_1\ra=|\psi_1\ra$
and that $\overline{T}$ commutes with $\Gamma$.
The operator $R\Pi$ is a linear combination of Pauli operators from the
group $\calS_1\cdot \calG$. One can easily check that this group
contains only stabilizers of the surface code and detectable errors.
Thus $\Gamma (R\Pi) \Gamma$ is proportional to $\Gamma$.
The latter has trivial action on the logical state $\overline{T} |\psi_1\ra$.
This proves that  $|\psi_8\ra=\overline{T} |\psi_1\ra$,
as claimed.

We make Algorithm~\ref{algo_1} partially fault-tolerant by
repeating syndrome measurements at step~2 and step~6
sufficiently many times.
The observed syndromes are used to compute an
error-corrected version of the phase $\sigma$ at step~4
and to perform the final error correction
after step~8. 
Accordingly, the algorithm can fail in two distinct
ways. First, the final state could be
$(\overline{T})^{-1} |\psi_1\ra$
instead of $\overline{T} |\psi_1\ra$
because some syndrome $\sigma_i$
has been flipped due to an error.
Such error 
results in 
a wrong phase $\sigma$ computed at step~4.
We shall refer to such events as
Pauli frame errors since they can be viewed as applying the $T$-gate in a wrong
Pauli frame.  Likewise, an undetected $X$ error that occurs
before step~5 on the qubit $q_{\mathrm{loc}}$
 would result in a Pauli frame error,
as can be seen from the identity
$\overline{T}_{\mathrm{loc}}X_1 =X_1(\overline{T}_{\mathrm{loc}})^{-1}$.
Secondly, the algorithm may 
fail if the final error correction
performed after step~8 
resulted in a logical error.
We would like to achieve an exponential suppression for
logical $X$ errors and, at the same time, ensure that 
Pauli frame and logical $Z$ errors occur with probability
at most $\kappa \varepsilon$, where $\varepsilon$ is the physical error rate and and $\kappa$ is a constant prefactor.  

To measure the syndrome of the surface code
$\calS_1$ we use a well-known
quantum circuit proposed in~\cite{dennis2002topological,fowler2009high}.
It requires one ancillary syndrome qubit
per each stabilizer, see Figure~\ref{fig:Kitaev}.
The circuit applies 
four rounds of CNOTs that compute the syndrome of each stabilizer into the respective
syndrome qubit. A
syndrome measurement cycle 
begins by resetting each syndrome qubit to $|0\ra$ or $|+\ra$ state for $Z$ and $X$ stabilizers respectively. The cycle ends
by measuring each syndrome qubit in the $Z$- or $X$-basis.
A circuit measuring the syndrome of $\calS_2$
requires only a few minor modifications,
see Figure~\ref{fig:Kitaev}.

For numerical simulations we chose the 
depolarizing noise model~\cite{fowler2009high}. 
It depends on a single error rate
parameter $\varepsilon$ such that each
operation in the circuit (a gate, a measurement, or a qubit reset)
becomes faulty with the probability $\varepsilon$.
Faults on different operations are independent. 
The model can be described by stochastic Pauli errors enabling efficient simulation using
the stabilizer formalism~\cite{aaronson2004improved}.
A faulty CNOT gate is modelled as the ideal CNOT followed by
a Pauli error drawn uniformly at random from the two-qubit Pauli group.
A faulty measurement is modelled as the ideal measurement whose
outcome is flipped. A faulty qubit reset prepares a basis state
orthogonal to the ideal one. 
A faulty idle qubit suffers from a Pauli error
$X$, $Y$, or $Z$.

\begin{figure}
\centerline{
\includegraphics[width=0.65\textwidth]{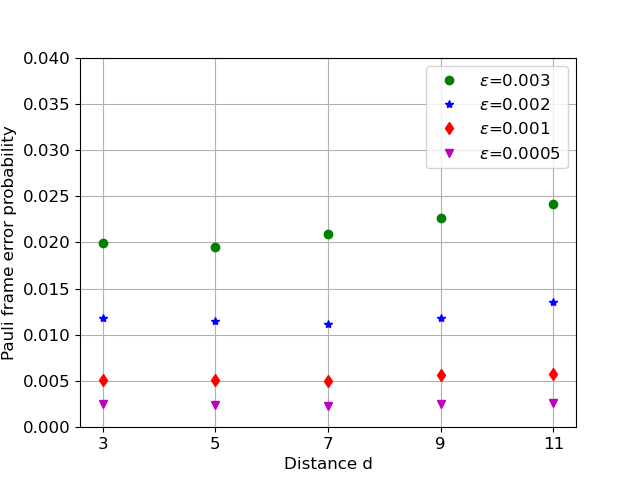}}
\caption{\textbf{Probability of a Pauli frame error} $P_F$ for the
code switching $\calS_1\to \calS_2$ 
with $L=3$ syndrome measurement cycles for the asymmetric
surface code $\calS_2$. The initial state is chosen
as an ideal logical state of the surface code $\calS_1$.
A Pauli frame error in the code switching
results in the implementation
of the logical gate $(\overline{T})^{-1}$ instead of $\overline{T}$. 
Our data suggest a scaling
$P_F\approx 6\varepsilon + O(d \varepsilon^2)$. 
\label{fig:plot1}
}
\end{figure}

\begin{figure}
\centerline{
\includegraphics[width=0.65\textwidth]{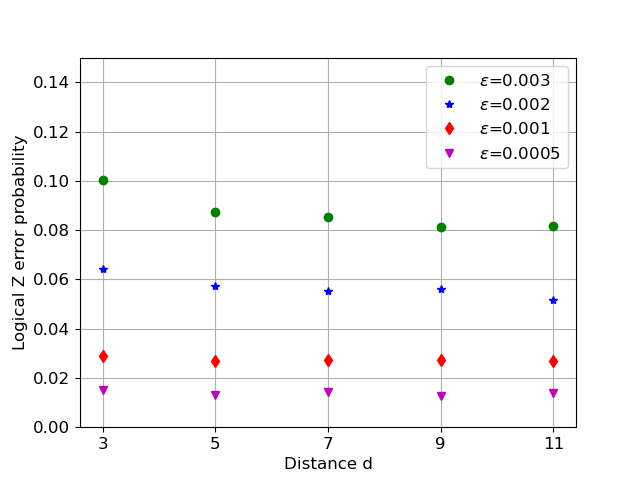}}
\caption{\textbf{Probability of a logical $Z$ error} 
$P_Z$
for the code switching $\calS_1\to \calS_2\to \calS_1$
with $L=3$ syndrome  cycles for the asymmetric
surface code $\calS_2$ and $d$ syndrome cycles for the
regular surface code $\calS_1$.
To enable efficient simulation we skipped the
$T$-gate application in Algorithm~\ref{algo_1}. 
Our data suggests a scaling
$P_Z\approx 26 \varepsilon$ 
for a large code distance.
The large constant prefactor stems from leaving the
logical qubit unprotected from $Z$ errors for $L+1$
syndrome cycles when the full syndrome information for
$X$ stabilizers of the code $\calS_1$ is unavailable.
\label{fig:plot2}
}
\end{figure}

Let us first discuss how to correct
Pauli frame errors.
Suppose $U$ is a quantum circuit describing
the first two steps of Algorithm~\ref{algo_1} with $L$ syndrome
measurement cycles at step~2. 
The circuit outputs a quantum state $|\psi_2\ra$
and a classical syndrome history $h$ specifying the syndrome of each stabilizer of $\calS_2$
measured in each of $L$ syndrome cycles. 
In the absence of errors all stabilizers except for
$G_1,\ldots,G_t$ have trivial syndromes.
Furthermore, the syndromes of $G_i$
measured in different cycles are the same. 
Let $\calF(U)$ be the set of all possible faulty implementations of $U$,
as prescribed by the depolarizing noise model.
Each circuit $\tilde{U}\in \calF(U)$ differs from $U$ by inserting Pauli errors
at some space-time locations (measurement and reset errors
can be modeled by inserting suitable Pauli errors before the ideal measurement
and after the ideal reset). 
We pick a faulty circuit $\tilde{U} \in \calF(U)$ randomly, according to the depolarizing noise model.
A simulation of the circuit $\tilde{U}$ yields a noisy syndrome history  
$\tilde{h}$.
To perform error correction we used 
the maximum-weight matching (MWM) decoder proposed in~\cite{bravyi2013simulation}
which is closely related to the one used in Refs.~\cite{dennis2002topological,fowler2009high}.
The MWM decoder
takes the syndrome history $\tilde{h}$ as an input
and outputs a candidate faulty circuit $\tilde{U}_{\mathrm{dec}}\in \calF(U)$
consistent with  $\tilde{h}$.
By propagating all Pauli errors contained in 
$\tilde{U}_{\mathrm{dec}}$ towards the final time step we 
compute error-corrected syndromes $\sigma_i$
and a Pauli correction $C$ acting on the data qubits
such that $CU|\psi_1\ra$ is our guess of the noisy
state $\tilde{U}|\psi_1\ra$, based on the available syndrome
information. Step~4 of Algorithm~\ref{algo_1} computes an error-corrected
phase $\sigma=(-1)^a \sigma_1 \cdots \sigma_t$,
where $a=1$ if $C$ contains an $X$ error
on the qubit $q_{\mathrm{loc}}$
and $a=0$ otherwise. To avoid the actual application
of the $T$-gate (which would require non-stabilizer
simulators) we chose the initial logical state
$|\psi_1\ra$ as an eigenvector of $\overline{Z}$,
that is, $|\psi_1\ra=|\overline{b}\ra$, where
$b\in \{0,1\}$ is a random bit.
Error correction succeeds if $\tilde{U}|\psi_1\ra$
is an eigenvector of $(\overline{T}_{\mathrm{loc}})^\sigma$
with the eigenvalue $(-1)^b$.
Otherwise, a Pauli frame error is declared.

Our numerical results for the probability of a Pauli frame
error $P_F$ are shown on  Figure~\ref{fig:plot1}.
Each data point represents an empirical
estimate of $P_F$ obtained by the Monte Carlo method
(we used between $50,000$ and $200,000$
Monte Carlo trials per data point). 
We chose $L=3$ syndrome measurement cycles to ensure that 
any single fault in the measurement of $\sigma_i$
is correctable.
Recall that the MWM decoder detects errors by examining 
differences between syndromes measured in subsequent cycles.
Thus $L=3$ provides two parity checks to diagnose errors
in the three measured values of $\sigma_i$,
in a way analogous to the $3$-bit repetition code.
In contrast, choosing $L=2$ would only allow to detect
a faulty bit $\sigma_i$ but not to correct it
(if the two measured values of $\sigma_i$ disagree, there
is no way to select the correct value). 
Choosing $L>3$ is undesirable as this leaves the logical qubit
unprotected from $Z$ errors for a longer time.
Our data suggests a scaling
$P_F\approx 6\varepsilon + O(d \varepsilon^2)$.
We numerically observed that the dominant contribution
to $P_F$ stems from
$X$ errors occurring on the qubit $q_{\mathrm{loc}}$
in the last syndrome cycle as well as syndrome measurement errors on the $Z$ stabilizer acting
on the qubit $q_{\mathrm{loc}}$ in the last syndrome cycle.
The term $O(d\varepsilon^2)$ can be understood as a result
of a weight-two error affecting the measurement of some 
syndrome bit $\sigma_i$. Here $i=1,\ldots,t$
can be arbitrary. Such weight-two error cannot be
corrected using only $L=3$ syndrome cycles. 

Finally, we estimated the probability of logical $X$ and $Z$ errors  by simulating a simplified version of Algorithm~\ref{algo_1} without steps~3,4,5. In other words, we skipped the
application of the physical $T$-gate. This enables
efficient simulation using the stabilizer formalism.
The MWM decoder takes as input the combined syndrome
history measured at steps~2,6 and calculates a Pauli correction
to be applied after step~8. 
The standard implementation
for MWM decoder~\cite{bravyi2013simulation}
was properly modified to
take into account that some stabilizers are turned off and on
during the code switching. 
Namely, syndromes of the operators
$G_i$ measured in the first cycle of step~2 as well as syndromes of the operators $F_i$ measured in the first cycle of
step~6 are not used to diagnose errors.  
As commonly done in the literature, we 
chose the number of syndrome cycles for the surface
code $\calS_1$ equal to the code distance $d$
and added a noiseless syndrome cycle at the end of the protocol.

\begin{figure}
\centerline{
\includegraphics[width=0.65\textwidth]{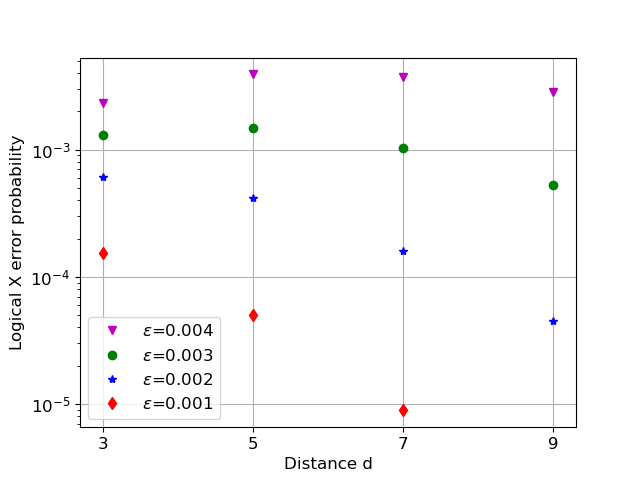}}
\caption{\textbf{Probability of a logical $X$ error} 
$P_X$
for the code switching $\calS_1\to \calS_2\to \calS_1$
with $L=3$ syndrome  cycles for the asymmetric
surface code $\calS_2$ and $d$ syndrome cycles for the
regular surface code $\calS_1$.
To enable efficient simulation we skipped the
$T$-gate application in Algorithm~\ref{algo_1}. 
Our data suggests that logical $X$ errors
are exponentially suppressed as one increases
the code distance.
\label{fig:plot2X}
}
\end{figure}

Our numerical results for the probability of 
logical $Z$ and $X$ errors $P_Z$ and $P_X$ are
shown on Figures~\ref{fig:plot2} and~\ref{fig:plot2X}, respectively.
Our data suggests a scaling
$P_{Z} \approx 26\varepsilon$ for a large
code distance. The large constant prefactor 
can be seen as the price we pay for
leaving the logical qubit unprotected from $Z$ errors during
$L$ syndrome cycles at step~2. In fact, since the syndromes
of $X$ stabilizers $F_i$ measured in the first cycle
of step~6 are random, these syndrome provide no information
about $Z$ errors that occurred in this cycle. 
Thus, the logical qubit is unprotected
from $Z$ errors for $L+1=4$ cycles. 
Since each cycle is a depth-$6$ circuit (four CNOT rounds,
reset round, and measurement round), the error correction
for $Z$ errors is effectively turned off for $24$ time steps.
This is in a good agreement with the scaling $P_{Z} \approx 26\varepsilon$.
Finally, we observed that the probability of logical $X$ errors is exponentially suppressed as one increases the code distance,
see Figure~\ref{fig:plot2X}.
Thus the noise in the implemented logical $T$-gate is dominated
by (coherent) $Z$ errors, as expected. 

We anticipate the scaling of $P_Z$ can be improved by optimizing the code switching protocol.
For example, one can reduce $P_Z$ roughly by the factor $2/3$
by choosing the number of syndrome measurement cycles $L$ for the code $\calS_2$ adaptively such that $L=2$ if the syndromes $\sigma_i$ measured in the first and the second cycles are the same for all $i=1,\ldots,t$ and $L=3$ otherwise. This would ensure that a single fault in the measurement of $\sigma_i$ can be corrected, similar to the $L=3$ implementation described above. However, the logical qubit remains unprotected from logical $Z$ errors for a shorter time (two cycles instead of three cycles, in the limit $\varepsilon\to 0$).

Under the assumption that we use a code with a sufficiently large distance, the logical $X$ errors are negligible. Hence the logical error rate $\bar \eps$ is determined by the logical $Z$ and Pauli frame errors.
The latter are described by the channel
$(1-P_F){\cal I} + P_F \calS^\dag$, where 
$\calS^\dag(B)=S^\dag BS$ for any single qubit operator $B$.
Twirling over the $X$ gate gives
a $Z$-type noise channel $(1-P_F/2){\cal I} + (P_F/2){\cal Z} $. Assuming that Pauli frame and logical $Z$ errors are independent, the combined logical error rate is
therefore $\bar{\varepsilon}\approx P_Z+P_F/2\approx 30\varepsilon$
in the limit $\varepsilon\ll 1$.

\section{Limitations and possible generalizations} \label{app_limitations}
With the usage of the QPD method in a fault-tolerant setting we inherit some of the shortcomings of this error mitigation scheme. The estimation of the logical error rate discussed in the main text needs to be sufficiently accurate and we require certain assumptions on the noise model, such as the absence of correlated errors between different realizations of error-mitigated gates and and the noise not strongly changing over time. The method is also restricted to removing the bias from expected values of measurement outcomes.
Furthermore, we require more accuracy on the transversal encoded gates, which is not a problem in practice. To see this, suppose for simplicity that the transversal operations are all executed without error, then the method would simulate a quantum channel \smash{$\mathcal{F}^{(\textnormal{ideal})}$} which is a quasi-probabilistic mixture \smash{$\cF^{(\textnormal{ideal})}=\sum_k a_k \cE^{(\textnormal{ideal})}_k$} of every 'branch' \smash{$\cE^{(\textnormal{ideal})}_k$} of the QPD. However, even transversal operations incur an error, which can be exponentially suppressed by increasing the code distance. Therefore the actual channel that is simulated by the method will be $\cF=\sum_k a_k \cE_k$ where $\cE_k$ includes the effects from noise in the transversal gates. We can make the error \smash{$\|\cE_k-\cE^{(\textnormal{ideal})}_k\| \leq \Delta$} arbitrarily small by increasing the code distance. By the triangle inequality we see that
\begin{align*}
    \norm{\cF - \cF^{(\textnormal{ideal})}} 
    \leq \sum_{k} |a_k| \norm{\cE_k-\cE^{(\textnormal{ideal})}_k} 
    \leq  \gamma_{\eps}^{2t} \Delta \, ,
\end{align*}
where $t$ denotes the $T$-count.
Hence to make the overall error $\delta>0$ on the output sufficiently small we have to choose $\Delta= \delta \gamma_{\eps}^{-2t}$ exponentially small in the number of $T$-gates. 
Recalling that in a generic fault-tolerant setting we choose 
\begin{align*}
 n
 =\polylog(1/\Delta, |C|) 
 =\polylog(\gamma_{\eps}^{2t}/\delta, |C|)
 =\poly\big(t,\log(1/\delta),\log(|C|)\big) \, ,
\end{align*}
which at first sight looks disappointing, because in realistic cases where $O(|C|)=O(t)$ the code size now scales polynomially in the number of locations $|C|$, instead of polylogarithmically as in the standard fault-tolerant setting. We argue that this is not really a problem in practice, since asymptotically the QPD method is in any case less advantageous than the conventional approach using a magic state factory. Recall that by nature of the QPD method, we only consider circuits where the total sampling overhead $\gamma_{\eps}^{2t}$ is sufficiently small, i.e., $\gamma_{\eps}^{2t} \leq \beta$ for some constant $\beta>0$ not too large~\footnote{If $\beta$ would be too large then the sampling overhead would make the QPD method infeasible.} we obtain $\Delta \geq \delta/\beta$ and hence
\begin{align}
    n 
    = \polylog(\beta/\delta,|C|) \, . \label{eq_practial_scaling}
\end{align}
The scaling~\eqref{eq_practial_scaling} matches again what we expect from the standard fault-tolerant setting.

In the near future one might encounter the situation where we lack the resources to distill magic states to sufficiently high fidelity that allows for large-scale fault-tolerant computation with $T$-gates, but a few rounds of distillation might still provide states that are better than raw noisy magic states. In that case one could combine distillation with error mitigation allowing for a trade-off between resources spent on distillation and the sampling overhead, by tuning how many distillation steps are performed.

\twocolumngrid
\bibliography{bibliofile}

\end{document}